\documentclass[12pt]{article}
\usepackage{amssym}
\usepackage{graphics}
\def\ap{a^{\dagger}}
\def\zz{\bar{z}}
\def\zzeta{\bar{\zeta}}
\def\xxi{\bar{\xi}}
\def\case#1#2{{\textstyle{#1\over #2}}}
\title{
\hfill{\normalsize ULB/229/CQ/01/2}\\
\vspace{1cm}
Completeness of photon-added squeezed vacuum and one-photon states and of
photon-added coherent states on a circle}

\author{C.\ Quesne \thanks{Directeur de recherches FNRS; E-mail:
cquesne@ulb.ac.be} \\ 
{\small \sl Physique Nucl\'eaire Th\'eorique et Physique
Math\'ematique,  Universit\'e Libre de Bruxelles,} \\ {\small \sl Campus de la
Plaine CP229, Boulevard~du Triomphe, B-1050 Brussels, Belgium}}
\date{ }
\begin{document}
\baselineskip=22pt plus 1pt minus 1pt
%%%%%%%%%%%%%%%%%%%%%%%%%%%%%%%%%%%%%%%%%%%%%%%%%%%%%%%%%%
\maketitle

\begin{abstract}
A discrete completeness relation and a continuous one  with a positive measure are found
for the photon-added squeezed vacuum states. Extension to the photon-added squeezed
one-photon states is considered. Photon-added  coherent states on a circle are introduced.
Their normalization and unity resolution relation are given.
\end{abstract}

\vspace{0.5cm}

\noindent
PACS: 03.65.Fd, 42.50.Dv

\noindent
Keywords: squeezed states, photon-added states, completeness relations

\bigskip\noindent
Corresponding author: C.\ Quesne, Physique Nucl\'eaire Th\'eorique et Physique
Math\'e\-ma\-ti\-que,  Universit\'e Libre de Bruxelles, Campus de la Plaine
CP229, Boulevard du Triomphe, B-1050 Brussels, Belgium

\noindent
Telephone: 32-2-6505559

\noindent
Fax: 32-2-6505045

\noindent
E-mail: cquesne@ulb.ac.be 
\newpage
%
%========================================================================
%
\section{Introduction}

Coherent states of the harmonic oscillator are known to have properties similar to those
of the classical radiation field~\cite{glauber, klauder, sudarshan}. There also exist states
of the electromagnetic field whose properties, such as squeezing~\cite{slusher},
higher-order squeezing~\cite{hong}, antibunching~\cite{kimble}, and sub-Poissonian
statistics~\cite{short}, are strictly quantum mechanical in nature. These states are called
nonclassical. Since the introduction of squeezed coherent states in the early
seventies~\cite{stoler}, many nonclassical states of the radiation field have been
constructed.\par
%
%----------------------------------------------------------------------------------------------------------
%
Among the latter, a class of states has attracted an ever increasing attention: the
so-called photon-added (or excited) states, which are obtained by repeated application of
photon creation operators on a given state and are distinct from displaced or squeezed
number states (for a review of the latter see e.g.~\cite{nieto}). The earliest example was
the photon-added coherent states (PACS), introduced by Agarwal and
Tara~\cite{agarwal91}. They were soon followed by the photon-added squeezed vacuum
state (PASVS), constructed by Zhang and Fan~\cite{zhang}. Since then photon-added
squeezed coherent states~\cite{xin}, photon-added thermal states~\cite{agarwal92},
photon-added even (PAECS) and odd (PAOCS) coherent states~\cite{dodonov}, for instance,
have been considered. Various methods of generating such states have also been
proposed~\cite{agarwal91, welsch}.\par
%
%---------------------------------------------------------------------------------------------------------
%
Some photon-added states have been interpreted in the context of nonlinear coherent
states related to a deformed oscillator~\cite{matos} or their
generalizations~\cite{mancini}. Such is the case for the PACS and the PASVS, which were
shown to be nonlinear coherent states~\cite{sivakumar} and even nonlinear coherent
states~\cite{liu}, respectively. Similarly, the photon-added squeezed one-photon states
(PASOPS) may be considered as odd nonlinear coherent states~\cite{liu}.\par
%
%--------------------------------------------------------------------------------------------------
%
Most of the theoretical studies of photon-added states have concentrated so far on
displaying their nonclassical properties, while leaving aside more fundamental questions,
such as their completeness. It should be stressed however that proving their
completeness is important both from theoretical and applied viewpoints. This property,
together with normalizability and continuity in the label, indeed makes them qualify as
generalized coherent states according to Klauder's prescription~\cite{klauder} on one
hand, and allows one to use them as (nonorthogonal) bases in many applications, on the
other hand.\par
%
%-----------------------------------------------------------------------------------------------------
%
The completeness of PACS has recently been proved by Sixdeniers and
Penson~\cite{sixdeniers01}. In the present letter, we consider the case of the PASVS and
PASOPS, as well as that of the photon-added coherent states on a circle (PACSC),
generalizing the PAECS and PAOCS of Ref.~\cite{dodonov}.\par
%
%=========================================================
%
\section{Definition and properties of photon-added squeezed vacuum states}

The PASVS are defined by~\cite{zhang}
\begin{equation}
  |\zeta, m\rangle = \left[N_m(|\zeta|)\right]^{-1/2} (\ap)^m |\zeta\rangle, 
  \label{eq:PASVS}
\end{equation}
where $m = 0$, 1, 2,~\ldots, $\ap$, $a$ are photon creation and annihilation operators
satisfying the relation $[a, \ap] = I$, $N_m(|\zeta|)$ is some normalization
coefficient, and 
\begin{equation}
  |\zeta\rangle = S(z) |0\rangle,   \label{eq:SVS}
\end{equation}
with $|0\rangle$ the vacuum state (i.e., $a |0\rangle = 0$). In~(\ref{eq:SVS}), $S(z)$ is
the squeezing operator
\begin{equation}
  S(z) = e^{\frac{1}{2} \left[z (\ap)^2 - \zz a^2\right]} = e^{\frac{1}{2} \zeta
  (\ap)^2} \left(1 - |\zeta|^2\right)^{\frac{1}{2} \left(N + \frac{1}{2}\right)} e^{-
  \frac{1}{2} \zzeta a^2}, 
\end{equation}
where $N = \ap a$ is the number operator and $\zeta$ is related to $z$ through the
relations
\begin{equation}
  z = r e^{{\rm i} \phi}, \qquad \zeta = \tanh r e^{{\rm i} \phi}. \label{eq:polar} 
\end{equation}
Hence, $\zeta$ is restricted to the unit disc ($|\zeta| < 1$) when $z$ runs over the
complex plane. In explicit form, the squeezed vacuum state~(\ref{eq:SVS}) can be
rewritten as
\begin{equation}
  |\zeta\rangle = \left(1 - |\zeta|^2\right)^{1/4} e^{\frac{1}{2} \zeta (\ap)^2}
  |0\rangle.  \label{eq:SVSbis} 
\end{equation}
In the limit $\zeta \to 0$ (resp.\ $m \to 0$), the state $|\zeta, m\rangle$ reduces to
the number state $|m\rangle = (m!)^{-1/2} (\ap)^m |0\rangle$ (resp.\ the squeezed
vacuum state $|\zeta\rangle$).\par
%
%-------------------------------------------------------------------------------------------------------
% 
{}From (\ref{eq:SVSbis}), it follows that the expansion of the states~(\ref{eq:PASVS}) in
the number-state basis $|n\rangle$, $n=0$, 1, 2,~\ldots, is given by
\begin{equation}
  |\zeta, m\rangle = \left[N_m(|\zeta|)\right]^{-1/2} \left(1 - |\zeta|^2\right)^{1/4}
  \sum_{k=0}^{\infty} \frac{\sqrt{(2k+m)!}}{k!}\, \left(\case{1}{2} \zeta\right)^k 
  |2k + m\rangle.  \label{eq:PASVS-exp}
\end{equation}
Hence, for a given $m$ value, the states $|\zeta, m\rangle$ belong to the subspace
${\cal F}^{(m)}_{\mu}$ of Fock space $\cal F$, spanned by the states $|2k +
m\rangle$, $k=0$, 1, 2,\ldots, with a photon number not less than $m$ and of the same
parity as $m = \mu\, {\rm mod} 2$.\par
%
%------------------------------------------------------------------------------------------------------
%
The overlap $\langle \xi, n | \zeta, m \rangle$ of two PASVS vanishes except if $|n-m|$ is
an even integer. If $n-m$ is a nonnegative even integer, the overlap can be written in any
one of the three following equivalent forms,
\begin{eqnarray}
  \langle \xi, n | \zeta, m \rangle & = & \left[N_m(|\zeta|) N_n(|\xi|)\right]^{-1/2}
       \left[\left(1 - |\zeta|^2\right) \left(1 - |\xi|^2\right)\right]^{1/4} 
       \frac{n!}{\left(\frac{n-m}{2}\right)!}\, (\case{1}{2} \zeta)^{(n-m)/2}
       \nonumber \\
  && \mbox{} \times {}_2F_1 \left(\frac{n+1}{2}, \frac{n+2}{2}; \frac{n-m}{2} + 1;
       \xxi \zeta\right) \nonumber \\
  & = & \left[N_m(|\zeta|) N_n(|\xi|)\right]^{-1/2} \langle \xi | \zeta \rangle
       \frac{n!}{\left(\frac{n-m}{2}\right)!}\, (\case{1}{2} \zeta)^{(n-m)/2}
       (1 - \xxi \zeta)^{-(n+m)/2}\nonumber \\
  && \mbox{} \times {}_2F_1 \left(- \frac{m-1}{2}, - \frac{m}{2}; \frac{n-m}{2} + 1;
       \xxi \zeta\right) \nonumber \\
  & = & \left[N_m(|\zeta|) N_n(|\xi|)\right]^{-1/2} \langle \xi | \zeta \rangle\, n!\, 
       \xxi^{(m-n)/4} \zeta^{(n-m)/4} (1 - \xxi \zeta)^{-(m+n)/4}\nonumber \\
  && \mbox{} \times P^{(m-n)/2}_{(m+n)/2} \left((1 - \xxi \zeta)^{-1/2}\right),  
       \label{eq:PASVS-overlap} 
\end{eqnarray}
where $\langle \xi | \zeta \rangle$ is the overlap of two squeezed vacuum states,
\begin{equation}
  \langle \xi | \zeta \rangle = \left[\left(1 - |\zeta|^2\right) \left(1 - |\xi|^2\right)
  \right]^{1/4} (1 - \xxi \zeta)^{-1/2}. 
\end{equation}
The first equality in~(\ref{eq:PASVS-overlap}) directly follows
from~(\ref{eq:PASVS-exp}) and the definition of the hypergeometric function
${}_2F_1(a, b; c; z)$, while the other two equalities result from well-known properties of
the latter and of Legendre functions of the first kind $P^{\mu}_{\nu}(z)$~\cite{erdelyi,
prudnikov}. If $n-m$ is a negative even integer, the corresponding overlap can be
deduced from~(\ref{eq:PASVS-overlap}) by using the Hermiticity property $\langle \xi, n |
\zeta, m \rangle = \overline{\langle \zeta, m | \xi, n \rangle}$.\par
%
%------------------------------------------------------------------------------------------------------
% 
As a special case of~(\ref{eq:PASVS-overlap}), we get back the overlap $\langle \zeta, n
| \zeta, m \rangle$ determined in~\cite{zhang} by a different method. We also obtain the
normalization coefficient of the PASVS,
\begin{equation}
  N_m(|\zeta|) = m!\, \left(1 - |\zeta|^2\right)^{-m/2} P_m \left((1 - |\zeta|^2)^{-1/2}
  \right),
\end{equation}
in terms of Legendre polynomials.\par
%
%-------------------------------------------------------------------------------------------------------
%
The states $|\zeta, m\rangle$ are distinct from the squeezed number states~\cite{nieto}
\begin{equation}
  |m, \zeta\rangle = S(z) |m\rangle, \qquad m = 0, 1, 2, \ldots.
\end{equation}
Contrary to the former, the latter are defined in a subspace of Fock space $\cal F$
including photon numbers less than $m$, namely the subspace ${\cal F}_{\mu}$ of even
or odd number states according to whether $m$ is even ($\mu = 0$) or $m$ is odd
($\mu = 1$). Since $S(z)$ is a unitary operator, the set of squeezed number states,
corresponding to a given $z$ or $\zeta$ value and $m = 0$, 1, 2,~\ldots, is an
orthogonal basis of $\cal F$:
\begin{eqnarray}
  \langle n, \zeta | m, \zeta \rangle & = & \delta_{n,m}, \\
  \sum_{m=0}^{\infty} |m, \zeta \rangle \langle m, \zeta | & = & I. \label{eq:SNS-RU}
\end{eqnarray}
\par
%
%----------------------------------------------------------------------------------------------------
%
Any PASVS can be expressed as a linear combination of squeezed number states
\begin{eqnarray}
  |\zeta, m\rangle & = & \left[(1- |\zeta|^2)^{m/2} P_m \left((1 -
         |\zeta|^2)^{-1/2}\right) \right]^{-1/2} \sqrt{m!} \nonumber\\
  && \mbox{} \times \sum_{k=0}^m \case{1}{2} \left[1 + (-1)^{m-k}\right]
         \frac{\zzeta^{(m-k)/2}}{(m-k)!! \sqrt{k!}}\, |k, \zeta \rangle, 
         \label{eq:PASVS-SNS}
\end{eqnarray}
and conversely
\begin{eqnarray}
  |m, \zeta \rangle & = & \sqrt{m!}\,  \sum_{k=0}^m \case{1}{2} \left[1 +
        (-1)^{m-k}\right] \left[(1- |\zeta|^2)^{k/2} P_k \left((1 - |\zeta|^2)^{-1/2} \right)
        \right]^{-1/2} \nonumber \\
  && \mbox{} \times  \frac{(- \zzeta)^{(m-k)/2}}{(m-k)!! \sqrt{k!}}\, |\zeta, k \rangle.
        \label{eq:SNS-PASVS}
\end{eqnarray}
In proving (\ref{eq:PASVS-SNS}), we used the property $S^{-1}(z) \ap S(z) = \left(\ap +
\zzeta a\right)/\sqrt{1 - |\zeta|^2}$, resulting from Baker-Campbell-Hausdorff formula,
and equation~(2.1) of~\cite{zhang}. We conclude that the set of PASVS corresponding
to a given $\zeta$ value and $m=0$, 1, 2,~\ldots forms a nonorthogonal basis of~$\cal
F$.\par
%
%===========================================================
%
\section{Completeness of photon-added squeezed vacuum states}

We may consider two different types of completeness or resolution of unity for the
PASVS: one in $\cal F$, obtained for a given $\zeta$ by summing over the discrete label
$m$, and the other in ${\cal F}^{(m)}_{\mu}$, obtained for a given $m$ by integrating
over the continuous label~$\zeta$.\par
%
%-------------------------------------------------------------------------------------------------------
%
The former directly follows from the unity resolution relation~(\ref{eq:SNS-RU}) for the
squeezed number states and the relation~(\ref{eq:SNS-PASVS}) between the latter and
the PASVS:
\begin{eqnarray}
  && (1 - |\zeta|^2)^{-1/2} \sum_{m=0}^{\infty} \sum_{n=m}^{\infty} (1 +
        \delta_{n,m})^{-1} \left[(n!/m!) P_m \left((1 - |\zeta|^2)^{-1/2}\right)
        P_n \left((1 - |\zeta|^2)^{-1/2}\right)\right]^{1/2} \nonumber \\
  && \times P^{(m-n)/2}_{(m+n)/2} \left((1 - |\zeta|^2)^{-1/2}\right)
        \Bigl[\left(- e^{-{\rm i} \phi}\right)^{(n-m)/2} |\zeta, m\rangle \langle \zeta, n|
        + \left(- e^{{\rm i} \phi}\right)^{(n-m)/2} |\zeta, n\rangle \langle \zeta,
        m|\Bigr] \nonumber \\
  && = I. 
\end{eqnarray} 
It has a nondiagonal form characteristic of a nonorthogonal basis, with coefficients given
by the elements of the overlap matrix inverse.\par
%
%---------------------------------------------------------------------------------------------------------
%
The derivation of the latter is more involved. The problem amounts to determining a
positive measure $d\rho_m(\zeta, \zzeta)$ such that
\begin{equation}
  \int d\rho_m(\zeta, \zzeta) |\zeta, m\rangle \langle \zeta, m| = I^{(m)}_{\mu},
  \label{eq:PASVS-RU}
\end{equation}
where the integration is carried out over the unit disc and $I^{(m)}_{\mu} \equiv
\sum_{k=0}^{\infty} |2k + m\rangle \langle 2k + m|$ denotes the unit operator in
${\cal F}^{(m)}_{\mu}$.\par
%
%-----------------------------------------------------------------------------------------------------
%
Making the polar decomposition $\zeta = |\zeta| e^{{\rm i} \phi}$, given
in~(\ref{eq:polar}), and the ansatz
\begin{eqnarray}
  d\rho_m(\zeta, \zzeta) & = & m!\, (1 - y)^{-(m+1)/2} P_m \left((1 - y)^{-1/2}\right)
       h_m(y) d^2\zeta, \nonumber \\ 
  y & \equiv & |\zeta|^2, \qquad d^2\zeta \equiv |\zeta| d|\zeta| d\phi,
\end{eqnarray}
and using the expansion~(\ref{eq:PASVS-exp}), we find after integrating over $\phi$
that equation~(\ref{eq:PASVS-RU}) is equivalent to the set of conditions
\begin{equation}
  \int_0^1 dy\, y^k h_m(y) = \frac{[(2k)!!]^2}{\pi (m+k)!}, \qquad k = 0, 1, 2, \ldots.
  \label{eq:moment}
\end{equation}
Consequently, the requirement that for a given $|m\rangle$, $|\zeta, m\rangle$ form a
complete (in fact, overcomplete) set in ${\cal F}^{(m)}_{\mu}$ is equivalent to the
resolution of a power-moment problem~\cite{akhiezer}.\par
%
%----------------------------------------------------------------------------------------------------
% 
As is usual in such a problem (see e.g.~\cite{sixdeniers01, sixdeniers99}), it is
convenient to set for complex $s$ and ${\rm Re}\, s > 0$, $k \to s-1$, to define
\begin{equation}
  g_m(y) = \left\{\begin{array}{ll}
      h_m(y) & {\rm if\ }0 < y < 1 \\[0.2cm]
      0 & {\rm if\ }1 < y < \infty
      \end{array}\right.,
\end{equation}
and to interpret (\ref{eq:moment}) as the Mellin transform $g^*_m(s)$ of
$g_m(y)$~\cite{sneddon},
\begin{equation}
  \int_0^\infty dy\, y^{s-1} g_m(y) = g^*_m(s) \equiv \left\{\begin{array}{ll}
      \frac{1}{2\pi} B\left(s, \frac{1}{2}\right) & {\rm if\ } m=1 \\[0.2cm]
      \frac{1}{4 \pi (m-2)!} B\left(s, \frac{m}{2}\right) B\left(s, \frac{m-1}{2}\right)
           & {\rm if\ } m = 2, 3, \ldots
  \end{array}\right.. \label{eq:mellin}
\end{equation}
In (\ref{eq:mellin}), $B(z,w)$ denotes the beta function, i.e., $B(z,w) = \Gamma(z)
\Gamma(w)/\Gamma(z+w)$. To find $g_m(y)$, we must perform the inverse Mellin
transform on $g^*_m(s)$.\par
%
%-------------------------------------------------------------------------------------------------------
%
{}For $m=1$, $g_1(y)$ is given in tables of Mellin transforms~\cite{prudnikov}, leading
to the following result for $h_1(y)$:
\begin{equation}
  h_1(y) = \frac{1}{2\pi} (1-y)^{-1/2}. \label{eq:h1}
\end{equation}
This is a positive function on (0, 1), increasing from $1/(2\pi)$ to $+\infty$ when y
goes from 0 to~1.\par
%
%---------------------------------------------------------------------------------------------------------
%
{}For higher $m$ values, it is convenient to use the Mellin convolution property of inverse
Mellin transforms, which states that if $g^*_m(s) = g^*_{m1}(s) g^*_{m2}(s)$ and
$g_{m1}(y)$, $g_{m2}(y)$ exist, then the inverse Mellin transform of $g^*_m(s)$ is
\begin{equation}
  g_m(y) = \int_0^{\infty} \frac{dt}{t}\, g_{m1}\left(\frac{y}{t}\right) g_{m2}(t).
  \label{eq:convolution}
\end{equation}
In applying (\ref{eq:convolution}) to (\ref{eq:mellin}) for $m \ge 2$, we choose
$g^*_{m1}(s) = [4\pi (m-2)!]^{-1} B (s, \frac{m}{2})$ and $g^*_{m2}(s) = B (s,
\frac{m-1}{2})$, which identifies $g_{m1}(y)$ and $g_{m2}(y)$ as 
\begin{equation}
  g_{m1}(y) = \left\{\begin{array}{ll}
      [4\pi (m-2)!]^{-1} (1-y)^{(m-2)/2} & {\rm if\ }0 < y < 1 \\[0.2cm]
      0 & {\rm if\ }1 < y < \infty
      \end{array}\right.,
\end{equation}
and
\begin{equation}
  g_{m2}(y) = \left\{\begin{array}{ll}
      (1-y)^{(m-3)/2} & {\rm if\ }0 < y < 1 \\[0.2cm]
      0 & {\rm if\ }1 < y < \infty
      \end{array}\right.,
\end{equation}
respectively~\cite{prudnikov}. Hence we obtain
\begin{equation}
  g_m(y) = [4\pi (m-2)!]^{-1} \int_y^1 dt\, t^{-m/2} (t-y)^{(m-2)/2} (1-t)^{(m-3)/2} 
  \label{eq:integral}
\end{equation}
for $0 < y < 1$, and $g_m(y) = 0$ for $1 < y < \infty$. It is obvious that the right-hand
side of~(\ref{eq:integral}) is a positive function, thus providing a solution for $h_m(y)$
for $m \ge 2$.\par
%
%-----------------------------------------------------------------------------------------------------------
% 
To express the latter in terms of known functions, we introduce a new variable $u =
(1-t)/(1-y)$, thereby obtaining
\begin{eqnarray}
  h_m(y) & = & [4\pi (m-2)!]^{-1} (1-y)^{m - \frac{3}{2}} \int_0^1 du\, u^{(m-3)/2} 
  (1-u)^{(m-2)/2} [1 - (1-y)u]^{-m/2}, \nonumber\\
  && m = 2, 3, \ldots. 
\end{eqnarray}
Formula 3.197.3 of~\cite{gradshteyn} then leads to
\begin{equation}
  h_m(y) = \frac{1}{2\pi (2m-3)!!}\, (1-y)^{m - \frac{3}{2}}\, {}_2F_1\left(\frac{m}{2},
  \frac{m-1}{2}; m - \frac{1}{2}; 1-y\right), \qquad m = 2, 3, \ldots.  
\end{equation}
By using formula 3.2.5 in volume~1 of~\cite{erdelyi}, we can rewrite $h_m(y)$, $m \ge
2$, in terms of a Legendre function of the second kind $Q^{\mu}_{\nu}(z)$, for which
$\mu = 0$ and $\nu$ is a nonnegative integer,
\begin{equation}
  h_m(y) = \frac{1}{2\pi (m-2)!}\, (1-y)^{(m - 2)/2} Q_{m-2} \left((1-y)^{-1/2}\right),
  \qquad m = 2, 3, \ldots. \label{eq:hm}  
\end{equation}
Such a function can be expressed in terms of Legendre polynomials combined with a
logarithmic function~\cite{erdelyi}:
\begin{eqnarray}
  Q_0 \left((1-y)^{-1/2}\right) & = & \frac{1}{2} \ln \frac{1 + \sqrt{1-y}}{1 -
        \sqrt{1-y}}, \label{eq:Q0}\\
  Q_{m-2} \left((1-y)^{-1/2}\right)  & = & \frac{1}{2} P_{m-2} \left((1-y)^{-1/2}\right)
        \ln \frac{1 + \sqrt{1-y}}{1 - \sqrt{1-y}} \nonumber \\
  && \mbox{} - \sum_{k=0}^{[(m-3)/2]} \frac{2m-4k-5}{(m-k-2) (2k+1)} P_{m-2k-3}
        \left((1-y)^{-1/2}\right), \nonumber \\
  && \mbox{} m = 3, 4, \ldots. \label{eq:Qm-2}  
\end{eqnarray}
Here $[x]$ denotes the largest integer contained in $x$.\par
%
%----------------------------------------------------------------------------------------------------------
%
{}For the first few values of $m$, we find
\begin{eqnarray}
  h_2(y) & = & \frac{1}{4\pi} \ln \frac{1 + \sqrt{1-y}}{1 - \sqrt{1-y}}, \\
  h_3(y) & = & \frac{1}{4\pi} \left( \ln \frac{1 + \sqrt{1-y}}{1 - \sqrt{1-y}} - 2
        \sqrt{1-y}\right), \\
  h_4(y) & = & \frac{1}{16\pi} \left[(2+y) \ln \frac{1 + \sqrt{1-y}}{1 - \sqrt{1-y}} - 6
        \sqrt{1-y}\right], \\
  h_5(y) & = & \frac{1}{144\pi} \left[3(2+3y) \ln \frac{1 + \sqrt{1-y}}{1 - \sqrt{1-y}} -
        2 (11+4y) \sqrt{1-y}\right].
\end{eqnarray}
\par
%
%-------------------------------------------------------------------------------------------------------
% 
{}From (\ref{eq:hm}), (\ref{eq:Q0}), and~(\ref{eq:Qm-2}), it can be shown that $h_m(y)
\to +\infty$ or 0 according to whether $y\to 0$ or~1. This is confirmed by Fig.~1, which
displays $h_m(y)$ for several $m$ values.\par
%
%-------------------------------------------------------------------------------------------------------
%
Having found a solution for the problem stated in~(\ref{eq:moment}), we may now ask
whether this solution is unique. An answer is provided by the (sufficient) condition of
Carleman~\cite{akhiezer}: if a solution exists and
\begin{equation}
  S \equiv \sum_{k=1}^{\infty} a_k, \qquad a_k \equiv \left(\frac{[(2k)!!]^2}{\pi
  (m+2k)!}\right)^{-1/(2k)},
\end{equation}
diverges, then the solution is unique. The convergence of~$S$ can be tested by applying
the logarithmic test~\cite{prudnikov}: if $\lim_{k \to \infty} [\ln(a_k) / \ln(k)] > -1$,
then $S$ diverges. By using Stirling formula for the asymptotic form of
$\Gamma(z)$~\cite{erdelyi}, we obtain $\lim_{k \to \infty} [\ln(a_k) / \ln(k)] = 0$.
We conclude that $h_m(y)$ given in~(\ref{eq:h1}) and~(\ref{eq:hm}) is the unique
solution to the problem.\par
%
%===========================================================
% 
\section{Extension to the photon-added squeezed one-photon states}

The PASOPS are defined by~\cite{liu}
\begin{equation}
  |1, \zeta, m\rangle = [N_{1m}(|\zeta|)]^{-1/2} (\ap)^m |1, \zeta\rangle,
\end{equation}
where $m = 0$, 1, 2,~\ldots, 
\begin{equation}
  |1, \zeta\rangle = S(z) |1\rangle = \left(1 - |\zeta|^2\right)^{3/4} e^{\frac{1}{2} \zeta
  (\ap)^2} |1\rangle,
\end{equation}
and $|1\rangle = \ap |0\rangle$. In the limit $\zeta \to 0$ (resp.\ $m\to 0$), they
reduce to the number state $|m+1\rangle$ (resp.\ the squeezed one-photon state $|1,
\zeta\rangle$).\par
%
%----------------------------------------------------------------------------------------------------------
%
Their expansion in the number-state basis is given by
\begin{equation}
  |1,\zeta, m\rangle = \left[N_{1m}(|\zeta|)\right]^{-1/2} \left(1 -
  |\zeta|^2\right)^{3/4}
  \sum_{k=0}^{\infty} \frac{\sqrt{(2k+m+1)!}}{k!}\, \left(\case{1}{2} \zeta\right)^k 
  |2k + m + 1\rangle,  
\end{equation}
showing that for a given $m$ value, they belong to the same subspace
${\cal F}^{(m+1)}_{\mu'}$ ($\mu' \equiv (m+1)\, {\rm mod} 2$) of $\cal F$ as the
PASVS $|\zeta, m+1\rangle$. We actually obtain
\begin{equation}
  [N_{1m}(|\zeta|)]^{1/2} \left(1 - |\zeta|^2\right)^{-1/2} |1,\zeta, m\rangle =
  [N_m(|\zeta|)]^{1/2} |\zeta, m+1\rangle,
\end{equation}
which enables us to easily extend some of the results of the two previous sections to the
PASOPS.\par
%
%---------------------------------------------------------------------------------------------------------
%
{}For instance, their overlap for $n-m$ an even nonnegative integer and their
normalization coefficient are given by
\begin{eqnarray}
  \langle 1, \xi, n | 1, \zeta, m \rangle & = & \left[N_{1m}(|\zeta|)
        N_{1n}(|\xi|)\right]^{-1/2} \langle 1, \xi | 1, \zeta \rangle\, (n+1)!\, 
        \xxi^{(m-n)/4} \zeta^{(n-m)/4} \nonumber \\
  && \mbox{} \times (1 - \xxi \zeta)^{-(m+n-2)/4} P^{(m-n)/2}_{(m+n+2)/2} \left((1 -
        \xxi \zeta)^{-1/2}\right), \label{eq:PASOPS-overlap}  
\end{eqnarray}
and
\begin{equation}
  N_{1m}(|\zeta|) = (m+1)! \left(1 - |\zeta|^2\right)^{-(m-1)/2} P_{m+1} \left((1 -
  |\zeta|^2)^{-1/2} \right),
\end{equation}
respectively. In (\ref{eq:PASOPS-overlap}), $\langle 1, \xi | 1, \zeta \rangle$ is the
overlap of two squeezed one-photon states,
\begin{equation}
  \langle 1, \xi | 1, \zeta \rangle = \left[(1 - |\zeta|^2) (1 - |\xi|^2)
  \right]^{3/4} (1 - \xxi \zeta)^{-3/2}. 
\end{equation}
\par
%
%------------------------------------------------------------------------------------------------------
%
Similarly, it can be shown that they form a nonorthogonal basis of ${\cal F}^{(1)}$ (i.e.,
the Fock space from which the one-dimensional subspace spanned by the vacuum state
has been removed) and an (over)complete set in ${\cal F}^{(m+1)}_{\mu'}$ with a
positive measure given by
\begin{eqnarray}
  d\rho_{1m}(\zeta, \zzeta) & = & (m+1)!\, (1 - y)^{-(m+2)/2} P_{m+1} \left((1 -
       y)^{-1/2}\right) h_{1m}(y) d^2\zeta, \qquad y \equiv |\zeta|^2, \nonumber \\
  h_{1m}(y) & = &\frac{1}{2\pi (m-1)!}\, (1-y)^{(m - 1)/2} Q_{m-1}
       \left((1-y)^{-1/2}\right), 
       \qquad m = 1, 2, \ldots. 
\end{eqnarray}
\par
%
%============================================================
%
\section{Definition and completeness of photon-added coherent states on a circle}

A special class of multiphoton coherent states is provided by the eigenstates of a power
$a^{\lambda}$ ($\lambda = 2$, 3, 4,~\ldots) of the photon annihilation
operator~\cite{buzek, sun, cq}, satisfying the relation
\begin{equation}
  a^{\lambda} |z, \mu\rangle = z |z, \mu\rangle, \qquad \mu = 0, 1, \ldots, \lambda-1.
  \label{eq:CSC-def}
\end{equation}
Here $\mu$ distinguishes between the $\lambda$ orthogonal solutions
of~(\ref{eq:CSC-def}), belonging to the subspaces ${\cal F}_{\mu}$ of Fock space $\cal
F$ spanned by the number states $|k \lambda + \mu\rangle$, $k=0$, 1, 2,~\ldots:
\begin{eqnarray}
  |z, \mu\rangle & = & [N_{\mu}(|z|)]^{-1/2} \sum_{k=0}^{\infty} \left(\frac{\mu!}
         {(k\lambda + \mu)!}\right)^{1/2} z^k |k\lambda + \mu\rangle,  \label{eq:CSC-exp1}\\
  N_{\mu}(|z|) & = & {}_0F_{\lambda-1} \left(\frac{1}{\lambda}+1, \frac{2}{\lambda}+1,
         \ldots, \frac{\mu}{\lambda}+1, \frac{\mu+1}{\lambda}, \frac{\mu+2}{\lambda},
         \ldots, \frac{\lambda-1}{\lambda}; y \right), \nonumber \\
  y & \equiv & |z|^2/\lambda^{\lambda}, 
\end{eqnarray}
where ${}_pF_q(a_1, \ldots, a_p; b_1, \ldots, b_q; z)$ denotes a generalized
hypergeometric function~\cite{erdelyi}.\par
%
%---------------------------------------------------------------------------
%
The states~(\ref{eq:CSC-exp1}) may also be written as linear combinations of $\lambda$
(standard) coherent states equidistantly separated from each other along a circle of
radius $|t| = |z|^{1/\lambda}$~\cite{sun},
\begin{eqnarray}
  |z, \mu\rangle & = & [N_{\mu}([z])]^{-1/2} \frac{\sqrt{\mu!}}{\lambda} e^{\frac{1}{2} |t|^2}
        t^{-\mu} \sum_{\nu=0}^{\lambda-1} \epsilon^{-\mu\nu} |t \epsilon^{\nu}\rangle,
        \qquad t^{\lambda} = z, \label{eq:CSC-exp2} \\
  |t \epsilon^{\nu}\rangle & = & e^{- \frac{1}{2} |t|^2 + t \epsilon^{\nu} \ap} |0\rangle, 
        \qquad \epsilon \equiv e^{2 \pi {\rm i}/\lambda}, \\
  N_{\mu}(|z|) & = & \mu!\, |t|^{-2\mu} h_{\mu+1}(|t|^2, \lambda), \label{eq:CSC-norm} 
\end{eqnarray}
hence the name of coherent states on a circle that is often used for them~\cite{janszky}.
In~(\ref{eq:CSC-norm}), $h_i(x,n)$ denotes a hyperbolic function of order~$n$, i.e., a
generalization of the hyperbolic cosine and sine functions to which it reduces for $n=2$
and $i=1$ or~2, respectively~\cite{erdelyi}.\par
%
%----------------------------------------------------------------------------
% 
Let us define PACSC by the relation
\begin{equation}
  |z, \mu, m\rangle = [N_{\mu m}(|z|)]^{-1/2} (\ap)^m |z, \mu\rangle,
\end{equation}
where $m=0$, 1, 2,~\ldots, $N_{\mu m}(|z|)$ is some normalization coefficient, and $|z,
\mu\rangle$ is given by~(\ref{eq:CSC-exp1}) or~(\ref{eq:CSC-exp2}). For $\lambda = 2$
and $\mu = 0$ or~1, they reduce to the PAECS or PAOCS, respectively~\cite{dodonov}.\par
%
%---------------------------------------------------------------------------
%
According to whether we use the expansions~(\ref{eq:CSC-exp1}) or~(\ref{eq:CSC-exp2}),
we can express the PACSC either in the number-state basis,
\begin{equation}
  |z, \mu, m\rangle = [N_{\mu m}(|z|) N_{\mu}(|z|)]^{-1/2} \sum_{k=0}^{\infty} \frac{[\mu!
  (k\lambda+m+\mu)!]^{1/2}}{(k\lambda+\mu)!} z^k |k\lambda + m + \mu\rangle,
  \label{eq:PACSC-def}
\end{equation}
or in terms of the PACS $|t \epsilon^{\nu}, m\rangle$ of~\cite{sixdeniers01},
\begin{equation}
  |z, \mu, m\rangle = [N_{\mu m}(|z|) N_{\mu}(|z|)]^{-1/2} [N_m(|t|)]^{1/2} \frac{\sqrt{\mu!}}
  {\lambda} e^{\frac{1}{2} |t|^2} t^{-\mu} \sum_{\nu=0}^{\lambda-1} \epsilon^{-\mu\nu} |t
  \epsilon^{\nu}, m\rangle,
\end{equation}
where
\begin{equation}
  |t\epsilon^{\nu}, m\rangle = [N_m(|t|)]^{-1/2} (\ap)^m |t \epsilon^{\nu}\rangle, \qquad
  N_m(|t|) = m!\, L_m(- |t|^2),  
\end{equation}
and $L_m(x)$ denotes a Laguerre polynomial~\cite{erdelyi}. From~(\ref{eq:PACSC-def}),
it is clear that for given $\mu$ and $m$ values, the states $|z, \mu, m\rangle$ belong to
the subspace ${\cal F}^{(m+\mu)}_{\mu_m}$ of Fock space $\cal F$ spanned by the states
with photon number $n$ not less than $m+\mu$ and congruent with $\mu_m$, defined by
$m + \mu = \mu_m {\rm mod} \lambda$.\par
%
%---------------------------------------------------------------------------
%
By using methods similar to those employed in Secs.~2 and~3, it is straightforward to
obtain the overlap of two PACSC and their normalization coefficient, as well as two
different kinds of completeness relations. Here we only mention two of these results.\par
%
%----------------------------------------------------------------------------
%
The normalization coefficient can be written either in closed form as 
\begin{eqnarray}
  N_{\mu m}(|z|) & = &  [N_{\mu}(|z|) \mu!]^{-1} (m+\mu)!\, {}_{\lambda}F_{2\lambda-1}
        \left(\frac{m+\mu+1}{\lambda}, \frac{m+\mu+2}{\lambda}, \ldots,
        \frac{m+\mu+\lambda}{\lambda}; \right. \nonumber \\
  && 1, \frac{1}{\lambda}+1, \frac{2}{\lambda}+1, \ldots, \frac{\mu}{\lambda}+1,
        \frac{\mu+1}{\lambda}, \frac{\mu+2}{\lambda}, \ldots, \frac{\lambda-1}{\lambda},
        \frac{1}{\lambda}+1, \frac{2}{\lambda}+1, \ldots, \nonumber \\
  && \frac{\mu}{\lambda}+1, \frac{\mu+1}{\lambda}, \frac{\mu+2}{\lambda}, \ldots, 
        \left.\frac{\lambda-1}{\lambda}; y\right), \qquad y \equiv
        |z|^2/\lambda^{\lambda},       
\end{eqnarray}
or as a linear combination of Laguerre polynomials,
\begin{equation}
  N_{\mu m}(|z|) = [\lambda N_{\mu}(|z|)]^{-1} \mu!\, m!\, |t|^{-2\mu}
  \sum_{\nu=0}^{\lambda-1} \epsilon^{-\mu\nu} e^{|t|^2 \epsilon^{\nu}} L_m(- |t|^2
  \epsilon^{\nu}).
\end{equation}
\par
%
%--------------------------------------------------------------------------
%
The PACSC satisfy a unity resolution relation in ${\cal F}^{(m+\mu)}_{\mu_m}$,
\begin{equation}
  \int d\rho_{\mu m}(z, \zz) |z, \mu, m\rangle \langle z, \mu, m| = I^{(m+\mu)}_{\mu_m},
\end{equation}
with a positive measure given by
\begin{eqnarray}
  d\rho_{\mu m}(z, \zz) & = & N_{\mu m}(|z|) N_{\mu}(|z|) (\mu!)^{-1} h_{\mu m}(y) d^2z,
         \qquad y \equiv |z|^2/\lambda^{\lambda}, \nonumber \\
  h_{\mu m}(y) & = & \frac{1}{\pi \lambda^{\lambda-\mu}} y^{(\mu+1-\lambda)/\lambda}
         e^{- \lambda y^{1/\lambda}} U\left(m, 1, \lambda y^{1/\lambda}\right),  
\end{eqnarray}
where $U(a, b, z) = \Psi(a, b; z)$ is Kummer's confluent hypergeometric
function~\cite{erdelyi}.\par
%
%===========================================================================
%
\section{Conclusion}

In the present letter, we demonstrated that the PASVS satisfy two different types of
unity resolution relations, a discrete one in $\cal F$ and a continuous one in ${\cal
F}^{(m)}_{\mu}$, and we extended such results to the PASOPS. In addition, we introduced
the PACSC and obtained both their normalization and their continuous unity
resolution relation in ${\cal F}^{(m+\mu)}_{\mu_m}$.\par
%
%-------------------------------------------------------------------------------------------------------
%
Proving the completeness of photon-added squeezed coherent states along similar lines is
a much more difficult problem since such states depend upon two continuous variables
instead of one~\cite{xin}. We hope however to solve it in a near future.\par
%
%--------------------------------------------------------------------------------------------------------
%
Another interesting open question is whether completeness relations of the second type
exist for photon-subtracted squeezed states. It is already clear that this is true neither
for photon-subtracted squeezed vacuum states, nor for the even nonlinear coherent
states proposed in~\cite{liu} by extending the results for positive integer $m$ values
to negative integer ones. In both cases, the states are indeed nonnormalizable in the limit
$\zeta \to 0$. Photon-subtracted squeezed excited states might, on the contrary, be
good candidates for the existence of completeness relations provided $m$ remains low
enough.\par
%
%----------------------------------------------------------------------------
%
As a final point, it is worth stressing that a central requirement of this work has been to
find a continuous resolution of unity of the usual type with a positive measure. Relaxing
this demand may lead to generalized unity resolution relations of the type considered in
Ref.~\cite{manko} for the nonclassical states studied in the present work, as well as for
their extensions.\par
%
%============================================================
% 

%
%========================================================================
% 
\newpage
\section*{Figure captions}

{}Fig.\ 1. The weight function $h_m(y)$ as a function of $y$ for various $m$ values: (a)
$m=1$ (solid line), $m=2$ (dashed line); (b) $m=3$ (solid line), $m=4$ (dashed line),
$m=5$ (dotted line).
\par
%
%======================================================================
% 
\newpage
\begin{picture}(160,100)
\put(35,0){\mbox{\scalebox{1.0}{\includegraphics{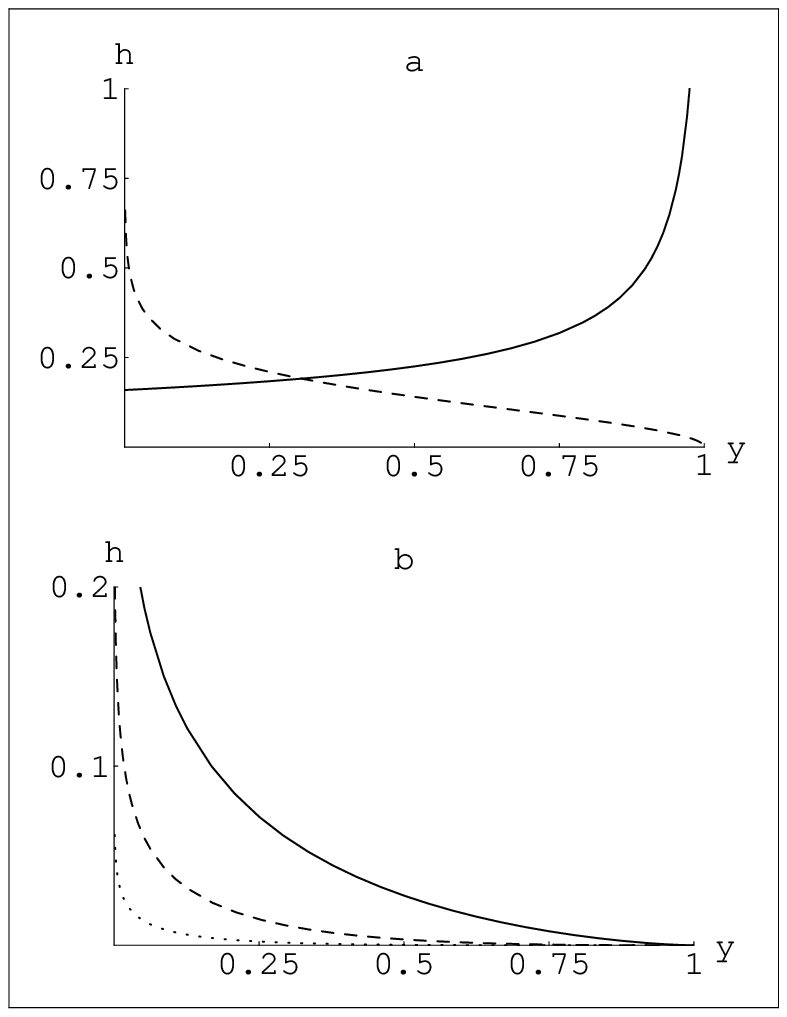}}}}
\end{picture}
\vspace{5cm}
\centerline{Figure 1}
%
%----------------------------------------------------------------------------------------------------------
%
%\newpage 
%\vspace*{17cm}
%\centerline{Figure 1}

\end{document}